\input cmtaco.sty
\font\fourteenbf=cmbx10 scaled \magstep2    
\magnification=\magstephalf

\vsize=9.0truein
\nohyphen
\mathfont\rm
\doublespace
\nopagenumbers
\headline={\ifnum\pageno>1\hss\tenrm --~\folio\tenrm ~-- \hss\else\hfil\fi}

\vskip 0.1cm 
\centerline{\fourteenbf A CEPHEID DISTANCE TO THE FORNAX CLUSTER}
\vskip 0.1cm 
\centerline{\fourteenbf AND THE LOCAL EXPANSION RATE OF THE UNIVERSE}
\vskip 1.5cm
\centerline{\bf BARRY F. MADORE$^1$, WENDY L. FREEDMAN$^2$, N. SILBERMANN$^3$}
\centerline{\bf PAUL HARDING$^4$, JOHN HUCHRA$^5$, JEREMY R. MOULD$^6$}
\centerline{\bf JOHN A. GRAHAM$^7$, LAURA FERRARESE$^8$}
\centerline{\bf BRAD K. GIBSON$^5$, MINGSHENG HAN$^9$, JOHN G. HOESSEL$^9$}
\centerline{\bf SHAUN M. HUGHES$^{10}$, GARTH D. ILLINGWORTH$^{11}$}
\centerline{\bf RANDY PHELPS$^2$, SHOKO SAKAI$^3$, PETER STETSON$^{12}$}

\vskip 0.8cm
\vfill
\singlespace
\noindent
--------------------------------

\par\noindent
$^1$  ~NASA/IPAC Extragalactic    Database,  Infrared  Processing  and
Analysis Center,

Jet Propulsion Laboratory, California Institute of Technology, MS~100-22,
Pasadena, CA ~91125
\par\noindent
$^2$ ~Observatories of the Carnegie Institution of Washington,  813  Santa    Barbara  St.,  

Pasadena, CA ~91101
\par\noindent
$^3$ ~Jet Propulsion Laboratory, California Institute of Technology,
MS~100-22, Pasadena, CA ~91125
\par\noindent
$^4$  ~Steward Observatory,  University of Arizona,  Tucson, AZ ~85721
\par\noindent
$^5$ ~Harvard-Smithsonian Institution,  Center   for Astrophysics,  60  Garden  Street,
Cambridge, MA ~02138
\par\noindent
$^6$ ~Mt. Stromlo and Siding Spring Observatories,  
Institute of Advanced Studies,  Private Bag, 

Weston Creek Post Office, ACT 2611,  Australia
\par\noindent
$^7$ ~Department  of Terrestrial   Magnetism, Carnegie  Institution  of
Washington, 

5241 Broad Branch Rd. N.W., Washington D.C. ~20015
\par\noindent
$^8$ ~Hubble  Fellow, California  Institute  of Technology,  MS~105-24
Robinson Lab, Pasadena, CA ~91125
\par\noindent
$^9$ ~Department of Astronomy, University of  Wisconsin, 475 N. Charter
St., Madison, WI ~53706
\par\noindent
$^{10}$ Royal Greenwich Observatory,  Madingley Road, Cambridge, UK ~CB3 OHA
\par\noindent
$^{11}$ Lick   Observatory, University of  California, Santa  Cruz, CA
~95064  
\par\noindent
$^{12}$  Dominion Astrophysical   Observatory,   5071 W.   Saanich   Rd.,
Victoria, BC, Canada ~V8X 4M6
\vfill\eject

\doublespace
\noindent

\par\noindent
{\bf The  Hubble  Space Telescope is being  used  to  measure accurate
Cepheid  distances   to  nearby galaxies  with   the  ultimate aim  of
determining the  Hubble constant,  H$_0$. For the   first time, it has
become feasible to  use Cepheid  variables to derive  a distance  to a
galaxy in  the  southern hemisphere cluster  of  Fornax. Based on  the
discovery of 37 Cepheids in the Fornax galaxy  NGC~1365, a distance to
this galaxy of  18.6$\pm$0.6~Mpc (statistical error only) is obtained.
This     distance  leads to  a   value    of  $H_0 =   70\pm7$(random)
$\pm18$(systematic)~km/sec/Mpc in good agreement with estimates of the
Hubble constant further afield.}

\par\noindent
The determination of an accurate value of the Hubble constant requires
the measurement of accurate galaxy distances and velocities.  Both the
Virgo and Fornax clusters of  galaxies are close  enough to be studied
in   considerable detail; on the   other  hand, they are  sufficiently
distant that the bulk of their measured velocity is probably dominated
by the  expansion of the universe.   The Fornax cluster of galaxies is
comparable  in distance  to the  Virgo  cluster$^1$,  but it  is found
almost opposite  to  Virgo in the  skies  of the southern  hemisphere.
Fornax is less rich in galaxies than  the Virgo cluster$^2$, but it is
also  substantially more compact  (see Figure~1).  As  a result of its
lower mass,  the influence of Fornax  on  the local velocity  field is
expectedly  less dramatic  than that  of  Virgo; and,  because  of its
compact nature, questions   concerning  the membership of   individual
galaxies    in Fornax  are less   problematic,   and the back-to-front
geometry  is far  less controversial  than in the   case  of the Virgo
cluster   complex$^3$.  Clearly, Fornax  is  an interesting site for a
direct test of  the local expansion rate if  its distance  were known.
With   Space Telescope, Cepheid   distances  out to  and including the
Fornax  cluster do offer  a direct  means of  estimating the expansion
rate of the nearby Universe.

\noindent
While  the goals of  the   Key Project on  the Extragalactic  Distance
Scale$^4$ are much broader than measuring  the distances to two nearby
clusters, there are several important  reasons to secure a distance to
the  Fornax  cluster.  Fornax serves   both  as a probe   of the local
expansion    velocity  field  and   as a   well   defined location for
calibrating several secondary distance indicators which can be used to
reach the essentially unperturbed Hubble flow.  To obtain the distance
to Fornax,  the Key Project is  searching for Cepheids in three member
galaxies.  The first of these, discussed here, is the luminous, barred
spiral galaxy, NGC~1365.

\noindent
At least three  lines of evidence  independently suggest that NGC~1365
is a physical member of the Fornax cluster.  First, NGC~1365 is almost
directly along our line of sight to Fornax as a whole, being projected
only $\sim$70 arcmin from the geometric center of the cluster, whereas
the  radius of  the cluster  is  $\sim$200~arcmin$^5$ (see  Figure~1).
Second,  NGC  1365 is also    coincident  with the  Fornax cluster  in
velocity space.   The  systemic (heliocentric) velocity  and  velocity
dispersion of the main population of galaxies in the inner six degrees
of   the  Fornax cluster    are  well defined  (Madore    et  al.   in
preparation):   39  spiral/irregular  galaxies  with published  radial
velocities give V = 1,399~km/sec and $\sigma = \pm334$~km/sec, 78~E/S0
galaxies give V = 1,463~km/sec with $\sigma =
\pm347$~km/sec; and  the combined (unweighted)  sample of 117 galaxies
gives    V  = 1,441~km/sec with $\sigma      = \pm342$~km/sec (see for
comparison 6,7).  We  note in passing  that the  mean velocity of  the
spirals agrees with the mean  for the ellipticals to within 64~km/sec.
And specifically, the  velocity difference  between NGC~1365 and   the
cluster  mean is only   about  two-thirds of the cluster's   intrinsic
velocity dispersion.  Third, and  finally, NGC~1365 sits only 0.02~mag
from  the central ridge line   of  the apparent Tully-Fisher  relation
relative to  other  cluster members   defined  by  recent  studies  of
Fornax$^{6,8}$.   Taken together  these observations  strongly suggest
that  a distance to  NGC~1365  alone should be  representative  of the
cluster as a whole.

\noindent
Using the  {\it Wide  Field and Planetary  Camera  2} on HST,  we have
obtained a  set   of  12-epoch   observations  of NGC~1365    and have
discovered 37 Cepheids  using two independent photometry routines, and
selected   by light-curve quality   (see  Silbermann {\it  et al.}  in
preparation, for details.)  The resulting intensity-weighted $<V>$ and
$<I>$   period-luminosity relations (Figure~2)  were fit  by $\chi ^2$
minimization  to  the  fiducial  PL  relation for  LMC Cepheids$^{9}$,
(corrected  for $E(B-V)_{LMC}  =$  0.10 mag) scaled    to an LMC  true
distance    modulus  of $\mu   _o$ =    18.50   mag, and  shifted into
registration with the Fornax data.   The resulting apparent moduli are
$\mu_V  =$  31.68$\pm$0.05~mag  and   $\mu_I  =$   31.55$\pm$0.05~mag.
Correcting for a total line-of-sight  reddening of $E(V-I)_{N1365} = $
0.14~mag (derived from    the Cepheid data  themselves,   and  using a
standard reddening law)   gives a true  distance modulus  of $\mu_0 =$
31.35$\pm$0.07~mag for NGC~1365.   This corresponds  to a distance  of
18.6$\pm$0.6~Mpc.  This error  quantifies  the statistical uncertainty
derived  from the  measured  photometric   errors, combined with   the
intrinsic magnitude and colour width of the Cepheid instability strip;
a more complete discussion of  the errors can be  found in Madore {\it
et  al.}   (in  preparation), giving  a    cumulative random error  of
$\pm$1.6~Mpc and   a systematic uncertainty  (quantifying  metallicity
effects, PL zero  point and photometric uncertainties)  also amounting
to $\pm$1.6~Mpc.

\par\noindent
Here, we  point out  that, unlike  our earlier discussion$^3$  of  the
expansion  rate derived from a Cepheid  distance to the galaxy M100 in
the Virgo cluster, the infall-velocity correction  for the Local Group
motion with respect   to the   Virgo   cluster  (and its    associated
uncertainty) becomes a minor issue  for the  Fornax cluster.  This  is
the result of a   fortuitous combination of geometry and   kinematics.
The physical separation of the  two clusters  is easily derived  since
both their angular separation and their individual distances are known
now known.  Under the assumption that the  Virgo cluster dominates the
local velocity perturbation field at the Local Group and at Fornax, we
can calculate the infall velocity   at Fornax (assuming that the  flow
field amplitude scales with $1/R_{Virgo}$, characterized by a $R^{-2}$
density  distribution$^{10}$).  From  this  we  then  derive  the flow
contribution to the  measured  line-of-sight radial velocity, as  seen
from the  Local Group.   Figure~3  shows the  distance scale  geometry
(left panel) and  the velocity-field kinematics  (right  panel) of the
Local Group--Virgo--Fornax  system.  Using an infall velocity$^{7}$ of
the Local Group toward Virgo  of +200~km/sec, and adopting a  generous
uncertainty of $\pm$100~km/sec, the flow correction for Fornax is only
--45 $\pm$23~km/sec. This correction and its uncertainty are more than
a factor of 4$\times$  smaller than the equivalent corrections applied
to M100.

\noindent
We  are now  in  a position to derive   two  estimates of  the general
expansion rate of   the local Universe.   The first  estimate is based
solely on the systemic  radial velocity and the Cepheid-based distance
to the Fornax cluster, and therefore samples the velocity field in one
particular direction    at  $\sim$20~Mpc.   We  then   examine a  more
isotropic (inner)  volume   of   space,  sampled  at    nine  randomly
distributed  distances  and directions out   to and including both the
Virgo and Fornax clusters.  This sample has the advantage of providing
an average over the volume, but it  is still limited in scale (probing
an average distance of $\sim$10 Mpc). It  is also still subject to the
uncertainties  in the  bulk flow  of  this  entire volume,  as well as
uncertainties in the adopted Virgocentric flow model.

\noindent
The observed velocity of the    Fornax cluster, 1,441~km/sec, can   be
corrected to  the  barycentre of the  Local   Group ($-$90~km/sec) and
compensating for the $-$45~km/sec  component of the Virgocentric flow,
derived  above,  to give a  cosmological expansion  rate  of Fornax of
$+1,306$~km/sec.   Using the  Cepheid distance of  18.6~Mpc for Fornax
then gives  $H_o  =  70~(\pm7)_r$~~[$\pm18]_s$ km/sec/Mpc.  The  first
uncertainty (in parentheses)  includes random  errors in the  distance
derived from  the PL fit  to  the  Cepheid  data,  as well  as  random
velocity  errors in the adopted   Virgocentric flow, combined with the
distance uncertainties to  Virgo  propagated through  the flow  model.
The second  uncertainty (in square  brackets) quantifies the currently
identifiable systematic  errors  associated   with  the  adopted  mean
velocity of  Fornax, and  the  adopted zero point  of the  PL relation
(combining in  quadrature the LMC   distance error,  a measure of  the
metallicity   uncertainty,   and a    conservative   estimate of   the
photometric  errors.  For an  extensive discussion of these errors see
the discussion and tabulation in Madore {\it et al.}  in preparation.)
Finally,  we note that   according  to the Han-Mould model$^{7}$,  the
so-called  ``Local Anomaly'' gives the   Local Group an extra velocity
component of approximately $+$73~km/sec towards Fornax.  If we were to
add  that correction our local estimate,  $H_o$ would increase from 70
to 74~km/sec/Mpc.

\noindent
Given the  highly   clumped nature  of  the   local  universe  and the
existence  of  large-scale streaming  velocities,   there is  still an
uncertainty remaining  due to the total peculiar  motion of the Fornax
cluster  with  respect to  the cosmic  microwave background restframe.
Observations of flows, and the determination of the absolute motion of
the  Milky Way with respect  to the  background radiation suggest that
line-of sight velocities  $\sim$300~km/sec  are not  uncommon.$^{11}$
The absolute motion  of Fornax with  respect to  the Local Group  then
becomes the  largest outstanding systematic  uncertainty at this point
in our  error  analysis: a 300~km/sec flow  velocity  for Fornax would
give a systematic error in the Hubble constant of $\pm$20\%.

We now   step back and  investigate   the Hubble flow between  us  and
Fornax, derived  from galaxies and  groups  of galaxies inside 20~Mpc,
each    having    Cepheid-based distances    and  expansion velocities
individually  corrected   for   a Virgocentric  flow  model$^{12,13}$.
Figure~4 presents these results in  graphical form.  Individual values
of $H_o$ range  from  61   to 99 km/sec/Mpc.    An average   of  these
independent determinations, including Virgo  and Fornax, gives $H_o =$
73~$(\pm4)_r~[\pm17]_s$~km/sec/Mpc.   This  determination,  as before,
uses a Virgocentric  flow model with  a  $1/R_{Virgo}$ infall velocity
fall-off,   scaled to a  Local  Group  infall velocity of +200~km/sec,
determined {\it ab initio}  by  minimizing the velocity  residuals for
the   galaxies     with   Cepheid-based   distances.    The  foregoing
determination of $H_o$ is again predicated on  the assumption that the
inflow-corrected velocities of both Fornax  and Virgo are not  further
perturbed by other mass concentrations or large-scale flows, such that
the 25,000 Mpc$^3$ volume of space delineated by  them is at rest with
respect to   the distant galaxy  frame.   An estimated flow correction
error  of $\pm300$~km/sec  dominates   the systematic  error   budget.
Nevertheless, these results are consistent  with those based on a more
extensive analysis of the greater flow field undertaken by Han {\it et
al.}  (in preparation).
\noindent

One of the interesting features of this  diagram is the relative small
residual scatter. Removing in  quadrature the 10\% scatter  introduced
by  the distance uncertainties leaves   an average velocity scatter of
less than  $\pm$70~km/sec about the  Hubble line.  That is, the random
velocities of nearby  galaxies, once corrected  for the the systematic
influence of    the Virgo  cluster,  are   remarkably small.   This is
consistent with earlier  discussions$^{14,15,16}$ that  call to attention
among other things the paucity of nearby blue-shifted galaxies outside
of the Local Group.

Further implications of a Cepheid distance  to the Fornax cluster, and
a comparison of the  global value of   the Hubble constant within  the
local expansion rate derived here can be  found in Madore {\it et al.}
(in  preparation).  But we note  that the estimates of H$_0$ presented
here  (based  on nearby   galaxies  out to and  including   the Fornax
cluster) are  in    close  agreement  with  the   value  of    H$_0  =
72$~km/sec/Mpc obtained from a variety of methods that operate at much
greater    distances.    These   methods  $^{17,18,19}$    include the
calibration  of type Ia and  type  II supernovae, and the Tully-Fisher
relation.

\centerline{\bf REFERENCES} 

\par\noindent
1. de~Vaucouleurs,  G., in {\it  Stars and Stellar Systems,}  {\bf 9},
(A.R.  Sandage, M.  Sandage, J.   Kristian, eds.), Univ Chicago Press,
557-596 (1975).

\par\noindent
2. Ferguson, H.C., \& Sandage, A.R.  Population Studies in Groups and
Clusters of Galaxies.  I.  The Luminosity Function  of Galaxies in the
Fornax Cluster.  {\it Astr.~J.}, {\bf 96}, 1520-1533 (1988).

\par\noindent
3.  Freedman, W.L., {\it et al.} Distance to the Virgo Cluster Galaxy
M100 from  Hubble Space Telescope  Observations of Cepheids.  {Nature}
{\bf 371}, 757-762 (1994).

\par\noindent
4. Kennicutt, R.C.,  Freedman,  W.L., \&  Mould, J.R.  Measuring the
Hubble Constant   with the Hubble  Space Telescope.  {\it Astron.~J.},
{\bf 110}, 1476-1491 (1995).

\par\noindent
5.  Ferguson, H.C.    Population  Studies in Groups  and   Clusters of
Galaxies. II. A Catalog of Galaxies in the Central  3.5 Degrees of the
Fornax Cluster. {\it Astr.~J.}, {\bf 98}, 367-418 (1989).

\par\noindent
6. Schroder, A.  UBVRI Photometry of Spiral Galaxies in the Virgo and
Fornax Clusters. ~Doctoral Thesis, University of Basel (1995).

\par\noindent
7.  Han  M.,  \& Mould  J.R.   The   Velocity  Field  in  the Local
Supercluster. {\it Astrophys.~J.,} {\bf 360}, 448-464 (1990).

\par\noindent
8.  Bureau,  M.,  Mould, J.R., \&  Staveley-Smith,  L.  A  New I-Band
Tully-Fisher Relation   for the  Fornax  Cluster: Implication  for the
Fornax   Distance   and   Local Supercluster  Velocity   Field.   {\it
Astrophys.J.,} {\bf 463}, 60-68 (1996).

\par\noindent
9.  Madore,    B.F., \& Freedman,    W.   L.  The   Cepheid  Distance
Scale. {\it Publs astron.  Soc.  Pacif.}, {\bf 103}, 933-957 (1991).

\par\noindent
10. Schechter, P. Mass-to-Light  Ratios for Elliptical Galaxies. {\it
Astron.~J.,} {\bf 85}, 801-811 (1980).

\par\noindent
11.  Coles,  P., \& Lucchin, F., in  {\it  Cosmology,} Wiley, 399-400
(1995)

\par\noindent
12. Kraan-Korteweg, R., A Catalog of 2810 Nearby Galaxies : The Effect
of the  Virgocentric  Flow Model on  their  Observed  Velocities. {\it
Astron.~Astrophys.~Suppl}, {\bf 66}, 255-279 (1986).

\par\noindent
13. Aaronson, M., Huchra, J., Mould, J., Schechter, P. \& Tully, R.~B.
The Velocity Field  in  the Local  Supercluster.  {\it Astrophys.~J.},
{\bf 258}, 64-76 (1980).

\par\noindent
14.  Sandage, A.R.  \&   Tammann,    G.A.  Steps Toward    the  Hubble
Constant.     V.  The Hubble Constant   from   Nearby Galaxies and the
Regularity of the  Local  Velocity Field.  {\it   Astrophys.~J.}  {\bf
196}, 313-328 (1975).

\par\noindent
15.  Sandage, A.R.,  Tammann, G.A. \&  Hardy, E.  Limits on  the Local
Deviation  of   the  Universe   from     a Homogeneous Model.     {\it
Astrophys.~J.}  {\bf 172}, 253-263 (1972).

\par\noindent
16. Fisher, J.R., \& Tully, R.B. Neutral  Hydrogen Observations of DDO
Dwarf Galaxies.  {\it Astron.~Astrophys.}, {\bf 44}, 151-171 (1975).

\par\noindent
17. Freedman, W.L., \& Mould, J.R.,  Kennicutt, R.C., \& Madore, B. F.
The Hubble Space Telescope Key Project to Measure the Hubble Constant.
IAU Symposium No. 183, ed. K. Sato, in press (1998).

\par\noindent
18. Reiss,  A., Press,  W.  \& Kirshner, R.   Using  Type Ia Supernova
Light Curve Shapes to Measure the Hubble Constant {\it Astrophys. J.,}
{\bf 438}, L17--L20 (1995).

\par\noindent
19.  Giovanelli. {\it  et al.}   The  Tully-Fisher Relation  and H$_0$
{\it Astrophys. J. Lett.,} {\bf 477}, pp. L1-L4 (1997)

\medskip
\noindent
{\bf ACKNOWLDGEMENTS. \  \ \ \ }  ~~This research was supported by the
National Aeronautics and Space Administration  (NASA) and the National
Science Foundation (NSF),  and  made extensive  use of  the  NASA/IPAC
Extragalactic Database (NED).  Observations are based on data obtained
using the {\it Hubble Space Telescope} which is  operated by the Space
Telescope   Science Institute under  contract from  the Association of
Universities  for Research in  Astronomy.  LF acknowledges support  by
NASA through  Hubble Fellowship grant  HF-01081.01-96A awarded  by the
Space   Telescope  Science  Institute,  which    is  operated  by  the
Association of Universities for Research in  Astronomy, Inc., for NASA
under contract NAS 5-26555

\medskip
Correspondence should be addressed to B.F.M (e-mail: barry@ipac.caltech.edu)

\vfill\eject
\centerline{\bf FIGURE CAPTIONS}
\par\noindent
{\bf Figure~1. --  \ \ } A  comparison of the distribution of galaxies
drawn to scale as  projected on the  sky for the Virgo  cluster (right
panel)  and the Fornax  cluster (left panel);  units  are arcmin.  The
comparison  of  apparent  sizes is   appropriate   given that  the two
clusters  are at  approximately  the same distance  from   us.  In the
extensive Virgo cluster, the galaxy M100 is marked $\sim$4\deg ~to the
north-west of the elliptical-galaxy-rich core;  this corresponds to an
impact parameter of 1.3~Mpc, or 8\% of the distance from the LG to the
Virgo  cluster.   The  Fornax  cluster    is  clearly  more  centrally
concentrated than   Virgo,  so  that the    back-to-front  uncertainty
associated with its   three-dimensional spatial extent is  reduced for
any randomly selected member.  Roughly  speaking, converting the total
angular extent of the Fornax cluster on the sky (6$\deg$ diameter$^5$)
into a  back-to-front extent, the  error associated with  any randomly
chosen galaxy in the  Fornax cluster,  translates  into a few  percent
uncertainty in  distance.  This uncertainty in  distance will  soon be
reduced when HST data for two additional  Fornax spirals (NGC~1425 and
NGC~1326A) are analyzed in the coming year.

\par\noindent
{\bf Figure~2.  -- \ \ } V  and I-band Period-Luminosity relations for
the full set of  37 high-quality Cepheids  monitored in NGC~1365.  The
fits are  to  the fiducial  relations$^{9}$ shifted   to the apparent
distance modulus  of NGC~1365.   Dashed  lines  indicate the  expected
intrinsic (2-sigma)   width of the   relationship  due  to the  finite
temperature width of the Cepheid instability strip.

\par\noindent
{\bf Figure~3.   --  \ \  } Relative  geometry (left panel),   and the
corresponding velocity vectors  (right panel) for  the disposition and
flow  of the Fornax cluster  and the Local  Group  with respect to the
Virgo cluster. The circles  plotted at the  positions of the Virgo and
Fornax clusters have  the same angular size  as the  circles minimally
enclosing M100 and NGC~1365 in the two panels of Figure~1.

\vfill\eject
\par\noindent
{\bf  Figure~4.   -- \ \ }   The velocity-distance relation  for local
galaxies   having  Cepheid-based  distances.   Circled dots   mark the
velocities  and  distances  of  the parent groups    or clusters.  The
one-sided     ``error'' bars  with   galaxy   names  attached mark the
velocities associated  with  the   individual galaxies having   direct
Cepheid distances. The thick broken line represents  a fit to the data
giving $H_o =  73$~km/sec/Mpc; the observed one-sigma (random) scatter
about  the mean  is  $\pm$12~km/sec/Mpc,  and is  shown   by the  thin
diverging broken   lines. All  velocities  have been  corrected  for a
Virgo-centric flow, as discussed in the text.

\vfill 
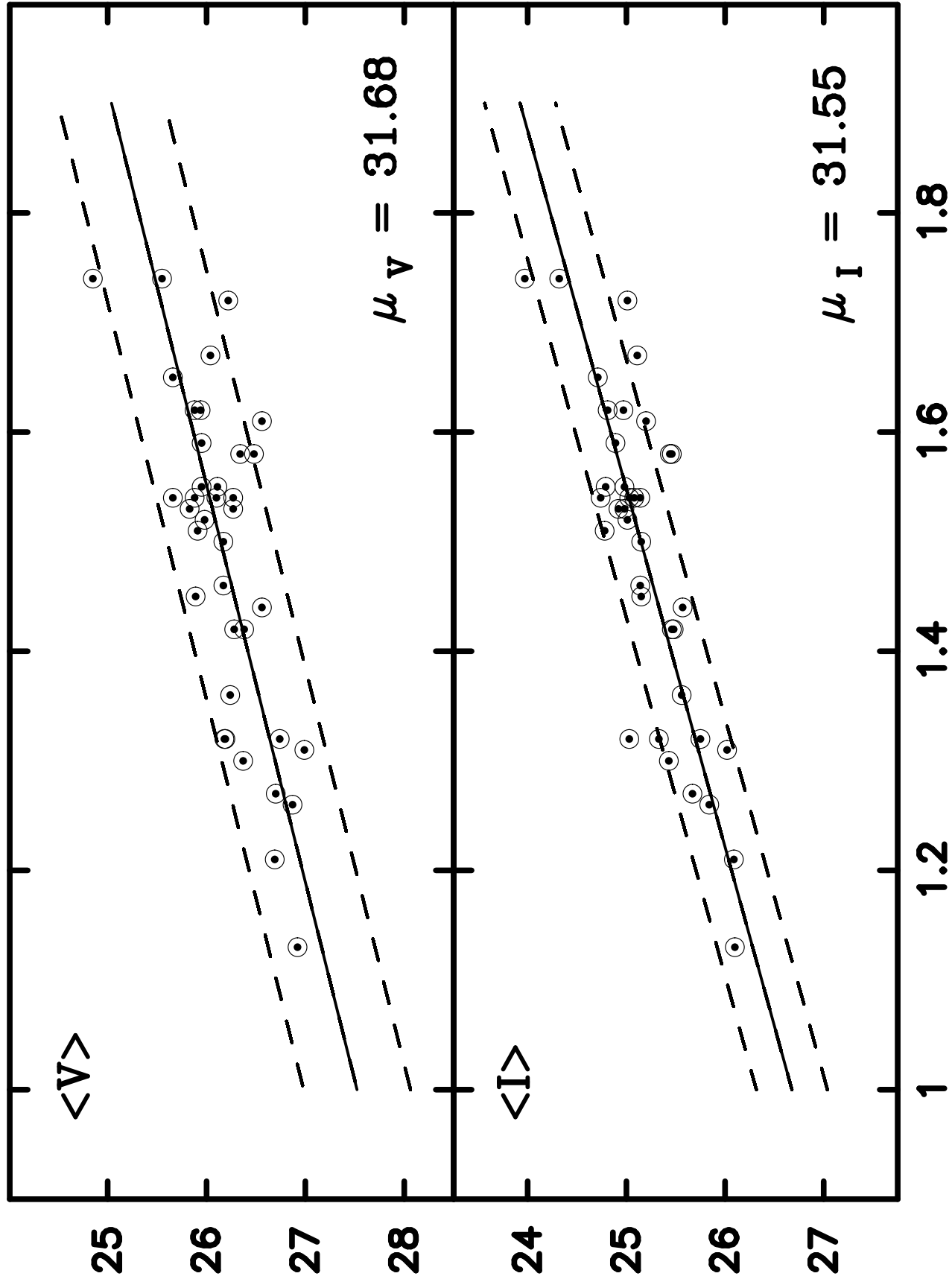
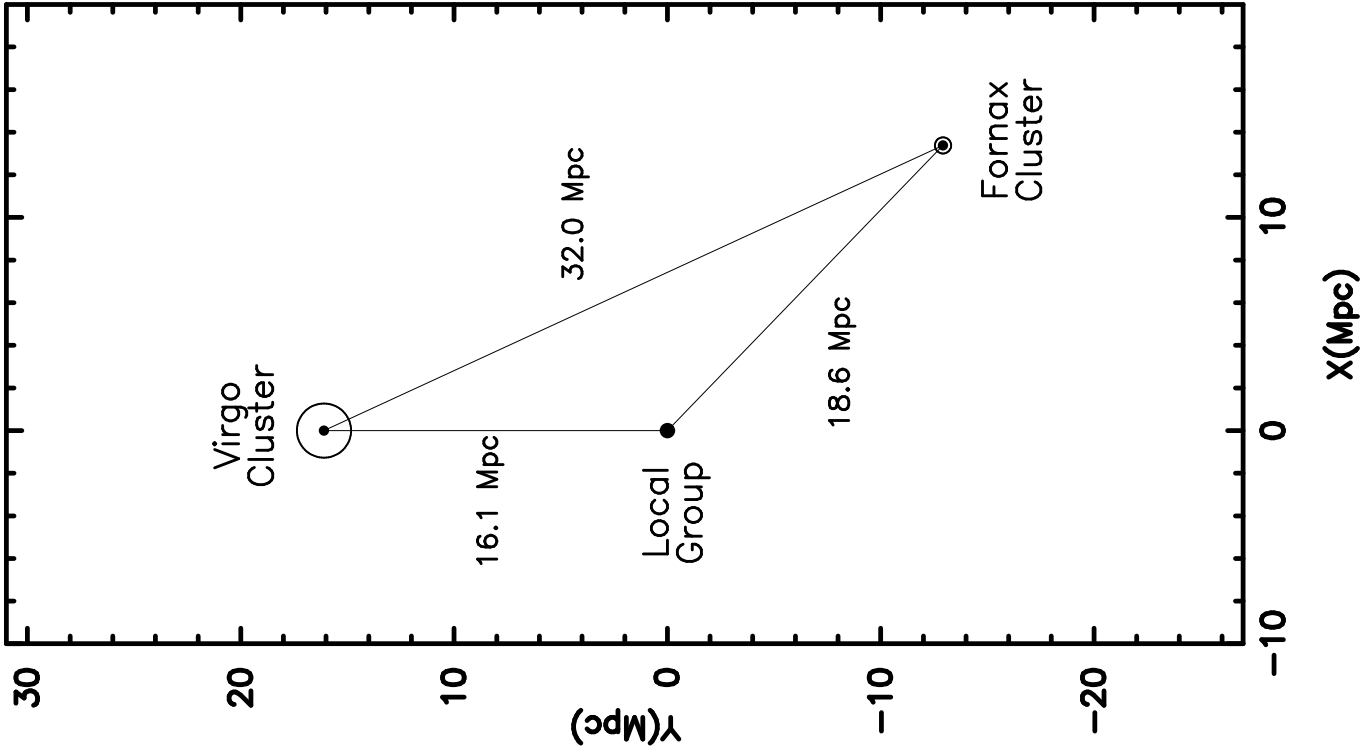
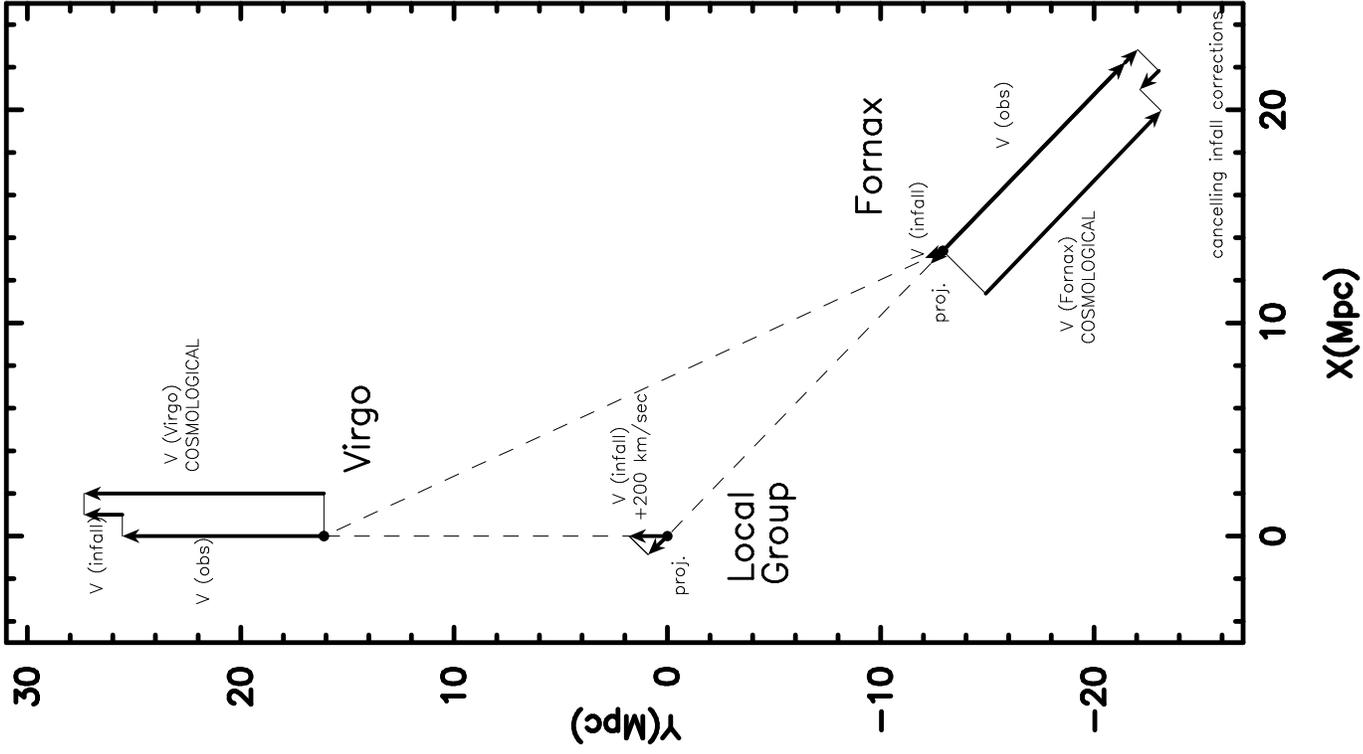
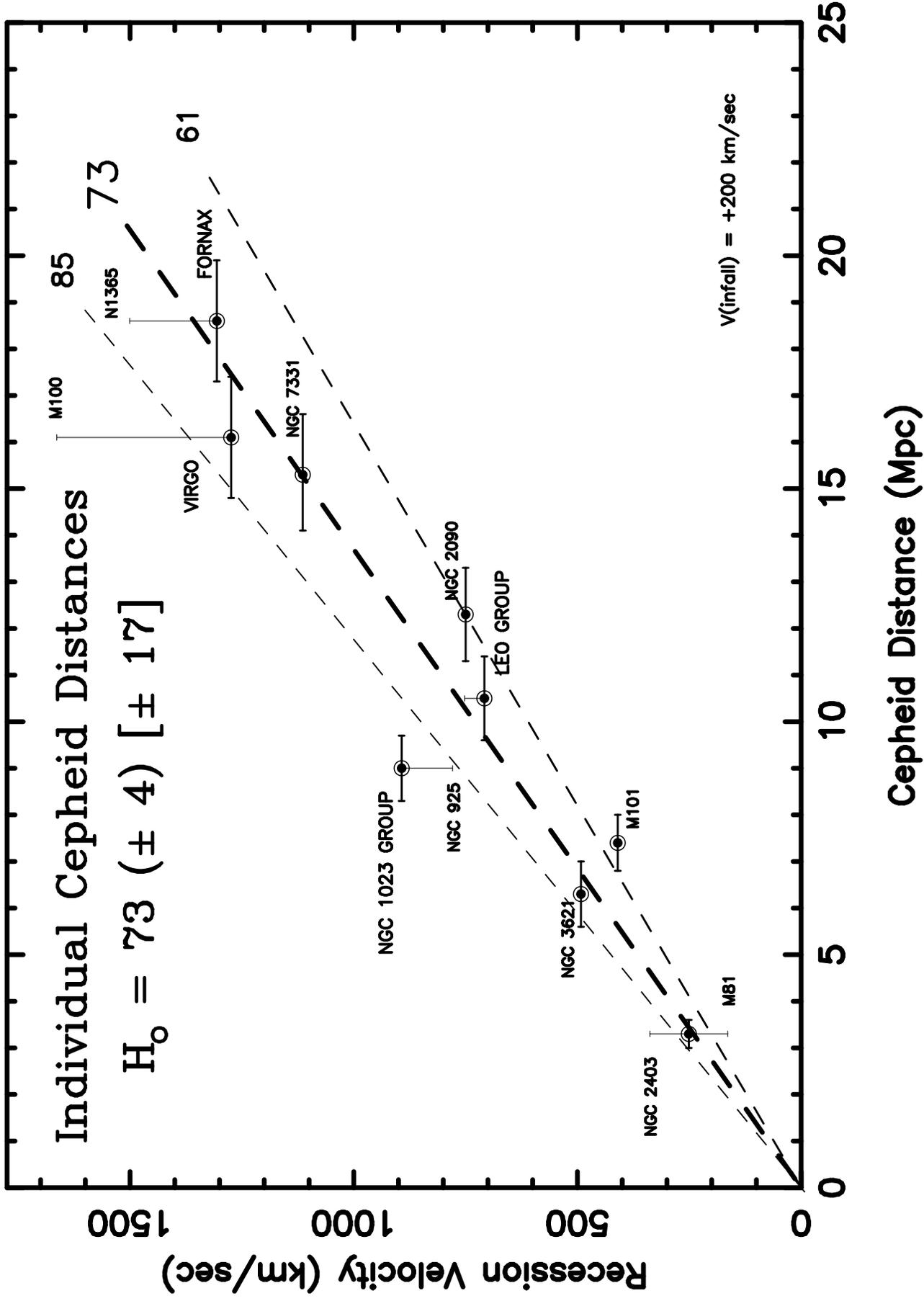
\bye